\documentclass[reprint,amsmath,amssymb,superscriptaddress,aps,prd]{revtex4-1}
\usepackage{placeins}
\usepackage{graphicx}
\usepackage{float}
\usepackage{amsmath}
\usepackage{appendix}
\usepackage{bm}
\usepackage[usenames, dvipsnames]{color}
\usepackage[utf8]{inputenc}

\newcommand{\NewChange}[1]{\textcolor{black}{#1}}

\begin{document}

\title{Transient glitch mitigation in Advanced LIGO data with $\textit{glitschen}$}
\author{J.D. Merritt}
\affiliation{Institute  for  Fundamental  Science, Department of Physics, University of Oregon, Eugene, OR 97403, USA}
\author{Ben Farr}
\affiliation{Institute  for  Fundamental  Science, Department of Physics, University of Oregon, Eugene, OR 97403, USA}
\author{Rachel Hur}
\affiliation{Institute  for  Fundamental  Science, Department of Physics, University of Oregon, Eugene, OR 97403, USA}
\author{Bruce Edelman}
\affiliation{Institute  for  Fundamental  Science, Department of Physics, University of Oregon, Eugene, OR 97403, USA}
\author{Zoheyr Doctor}
\affiliation{Institute  for  Fundamental  Science, Department of Physics, University of Oregon, Eugene, OR 97403, USA}
\affiliation{Center for Interdisciplinary Exploration and Research in Astrophysics (CIERA), Department of Physics and Astronomy, Northwestern
University, Evanston, IL 60201, USA}
\date{\today}

\begin{abstract}

``Glitches'' -- transient noise artifacts in the data collected by gravitational wave interferometers like LIGO and Virgo -- are an ever-present obstacle for the search and characterization of gravitational wave signals. With some having morphology similar to high mass, high mass-ratio, and extreme-spin binary black hole events, they limit sensitivity to such sources. They can also act as a contaminant for all sources, requiring targeted mitigation before astrophysical inferences can be made. We propose a data driven, parametric model for frequently encountered glitch types using probabilistic principal component analysis. As a noise analog of parameterized gravitational wave signal models, it can be easily incorporated into existing search and detector characterization techniques. We have implemented our approach with the open source \textit{glitschen} package. Using LIGO's currently most problematic glitch types, the `blip' and `tomte', we demonstrate that parametric models of modest dimension can be constructed and used for effective mitigation in both frequentist and Bayesian analyses.

\end{abstract}
\maketitle

\section{Introduction}

Detecting gravitational waves (GWs) is an immense challenge, requiring the construction and monitoring of the most sensitive interferometers ever built \cite{aligo_detectors}. The strain signal from a loud binary black hole (BBH) inspiral typically perturbs the detectors' arm lengths to one part in $10^{21}$. Managing the noise background is an overwhelming portion of that challenge: an earthquake in another hemisphere, a passing vehicle, a cosmic ray hit, a thirsty raven \cite{ravensAlog}, or scattered light from a blinking LED can all bring the data well short of the level necessary for detection \cite{detchar}. In spite of a myriad of obstacles, the LIGO-Virgo Collaboration (LVC) has detected 58 confident compact binary coalescence (CBC) events as of the end of the first half of the third observing run (O3a) \cite{gwtc2, gwtc21}. Into O4 these observatories may see confident CBC signals upwards of once-a-day \cite{prospectso3o4o5}. Upgrades to the detectors will improve sensitivity and the addition of KAGRA to the LIGO-Virgo-KAGRA network (LVK) will improve astrophysical \NewChange{parameter estimation (PE)} and sky localization. The LVK still expects serious challenges overcoming the noise background, carefully examining more near-threshold triggers, and keeping all the pipelines going with the rapid acquisition of a larger volume of data.

In Advanced LIGO data there are some transient noise sources for which no physical cause has been identified \cite{blips2019}. These noise sources have the potential to impact astrophysical searches significantly \cite{mullerGravitationalwaveAstronomyCompact2018}. In particular, high-mass and high-mass-ratio BBH searches are affected, in which the astrophysical hypothesis predicts a short duration signal sweeping up into the sensitive frequency bands of the detectors near merger. ``Blip" glitches and the lower frequency, longer duration ``tomte" glitches are glitch types that are capable of masquerading as these high mass CBCs. These occur on the order of 1/hour \cite{blips2019} but sometimes much more frequently, so the probability of coincidence in multiple detectors is non-negligible. Coincident or nearly coincident glitches can confuse search pipelines that strongly rely on coherence between detectors to determine if a trigger is astrophysical. Worse, the effect on the ranking statistic, established by time-sliding data streams from multiple detectors to establish false alarm rates (FARs) \cite{noise_guide} is affected significantly by the presence of these glitches in the background, effectively down-ranking many events. There is evidence that these glitches grow louder and more prevalent with increasing sensitivity \cite{gspy2021}. Blips and tomtes all but eliminate our ability to evaluate high-mass, extreme-mass-ratio, and extreme-spin single-detector triggers \cite{Davis_2020} from a confident astrophysical perspective because the data is contaminated with $\mathcal{O}(10^{4})$ loud glitches. 

GravitySpy \cite{gspy1} is a pipeline developed to classify glitch types. It leverages citizen science with an image recognition neural network, specifically trained on q-transforms, which display power in time-frequency pixels \cite{q-transf}. Thanks to these efforts, there are now over $10^{6}$ glitches classified, each with an associated confidence metric and SNR \cite{gspy1}. GravitySpy itself can be used to effectively distinguish different types of glitches from each other, but it can't be used to distinguish signal from glitch, or to subtract glitches from data. For this we seek a parametric, \emph{generative} model for common glitch types. Barring the discovery and mitigation of possible environmental, electronic, or instrumental causes \cite{Braginsky_2006, Yamamoto_2008, AHC_blips_slides} for these problematic classes of glitch, distinction between glitch-like astrophysical events and BBH signals that resemble common and problematic glitch types may be our only tractable method for opening up the high-mass and high-mass-ratio region of CBC search parameter space. 

With the \textit{glitschen} package, we propose a data-driven, easy to use, and computationally cheap framework for the modeling of short duration transient glitches. Our model uses an analytical maximum likelihood estimation (MLE) approach to fit a probabilistic principal component analysis (PPCA) model to all of the training data, operating under the hypothesis of a transient glitch superimposed on Gaussian noise~\cite{microsoft_PPCA}. While PCA's have previously been used in the context of glitch categorization~\cite{PCAT1,PCAT2}, we focus on the construction of glitch-class-specific parameterized models for glitch \emph{mitigation}.  Relative to other glitch mitigation techniques~\cite{Mukherjee_2010, bayeswave,dictionary_learning,Cornish:2020dwh_glitchbuster,Cornish:2021wxy_rapidPE}, these targeted parameterized models have minimal flexibility and are in many ways analogous to the parameterized CBC models used to search for and characterize signals, making them straightforward to incorporate in existing LVK analyses. 
\NewChange{In comparison to current glitch mitigation techniques such as \textsc{BayesWave} (BW) \cite{bayeswave}, our approach naturally allows for informed priors, allowing us to leverage the extensive glitch population. Our approach can be naturally used in existing analysis libraries such as \textsc{Bilby}, whereas BW's use of reverse-jump sampling means that only a point estimate from BW can be used to remove a glitch during astrophysical parameter inference. While powerful, our methods require large training sets for each glitch type and will likely be unable to model glitches that are extensive in time-frequency, such as scattered-light.}

\section{Method}
\subsection{Modeling the Advanced LIGO Noise Background}

The noise in the detector is a superposition of many noise sources, and is modeled as a stochastic process, drawing randomly from a stationary background spectrum at each frequency \cite{creightonAnderson}. The detector produces a time series, $n(t)$, which we can represent as a vector, $\bm{n}$. Transforming to the frequency domain we obtain $\bm{\tilde{n}}$, with $n_{i}$ indicating the noise in the i-th frequency bin. Assuming Gaussianity the probability distribution becomes: 
\begin{equation}
p(\bm{\tilde{n}}) = \frac{1}{det(2\pi\bm{C})^{1/2}} exp \bigg{[} - \frac{1}{2} \sum_{ij} (\tilde{n_{i}} - \mu) (\tilde{n_{j}} - \mu)C_{ij}^{-1}\bigg{]}
\end{equation}
where $\bm{C}_{ij} = \frac{1}{M-1}(n_{i} - \mu)(n_{j}-\mu)$ is the covariance matrix of the observations and $\mu$ is the mean of the data\cite{noise_guide}. Stationarity means that the noise spectrum is not changing over time, so in the frequency domain the covariance matrix is diagonal: $C_{ij}=\delta_{ij}S_{n}(f_{i})$, giving the power spectral density (PSD), $S_{n}(f)$ which is equal to the square of the amplitude spectral density (ASD). The noise is typically stationary on the timescales (minutes) relevant for PSD computation, but on the hour timescale may need to be updated \cite{noise_guide}.

We ``whiten" the data by dividing the frequency domain data by an estimate of the ASD, resulting in noise with an equal (unitary) noise in all frequencies. We train and test our model using whitened data.

This treatment is highly effective for `well-behaved' noise sources which remain stationary over the duration of ASD calculation, however the motivation for building our model is to mitigate transient glitches, which can occur at any time and pose the greatest challenge for searches that look for transient astrophysical events.

\begin{figure}

\begin{minipage}{0.4\textwidth}
    \includegraphics[width=\linewidth]{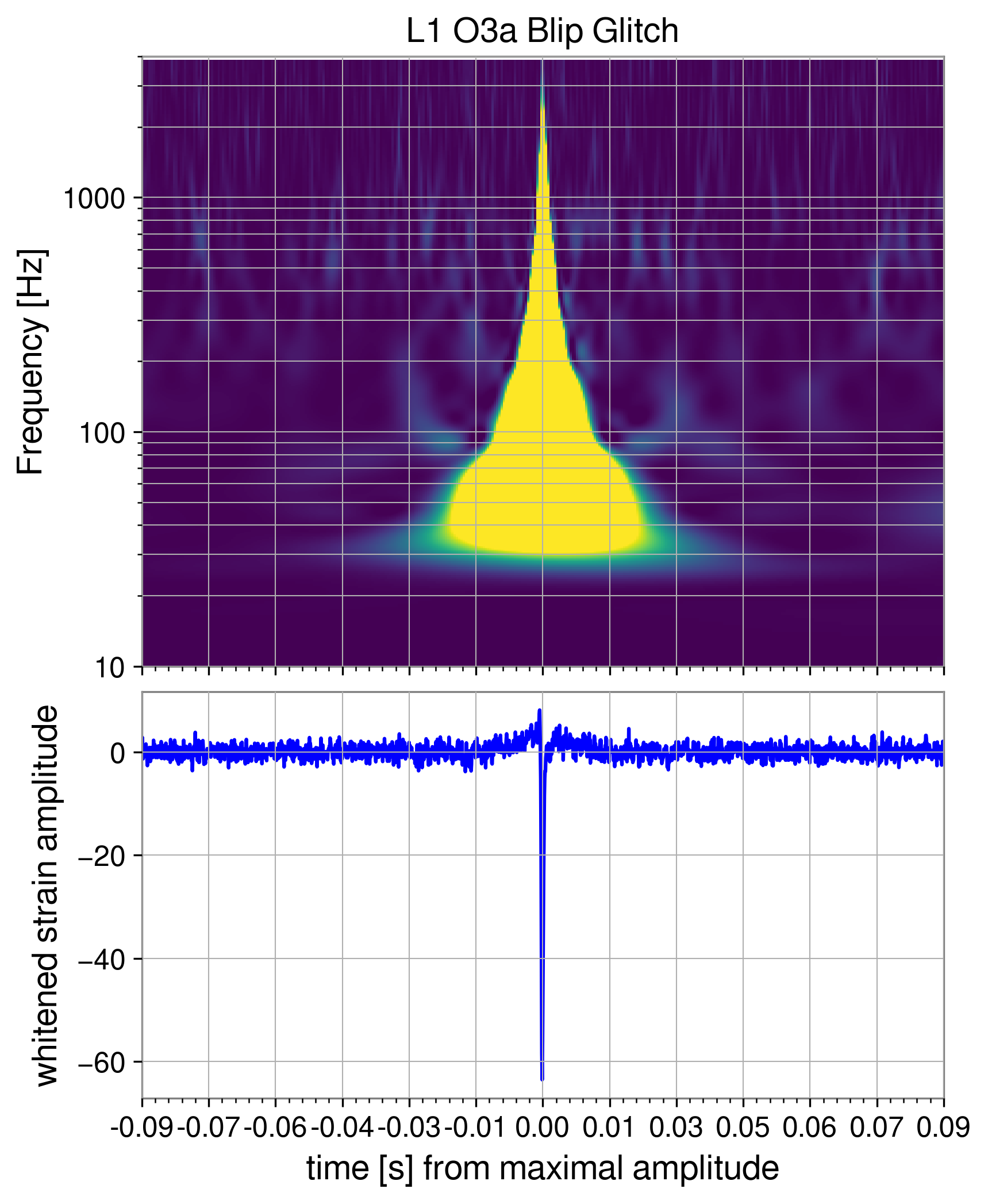}
    \end{minipage}
    \hspace{\fill} 
    \begin{minipage}{0.4\textwidth}
    \includegraphics[width=\linewidth]{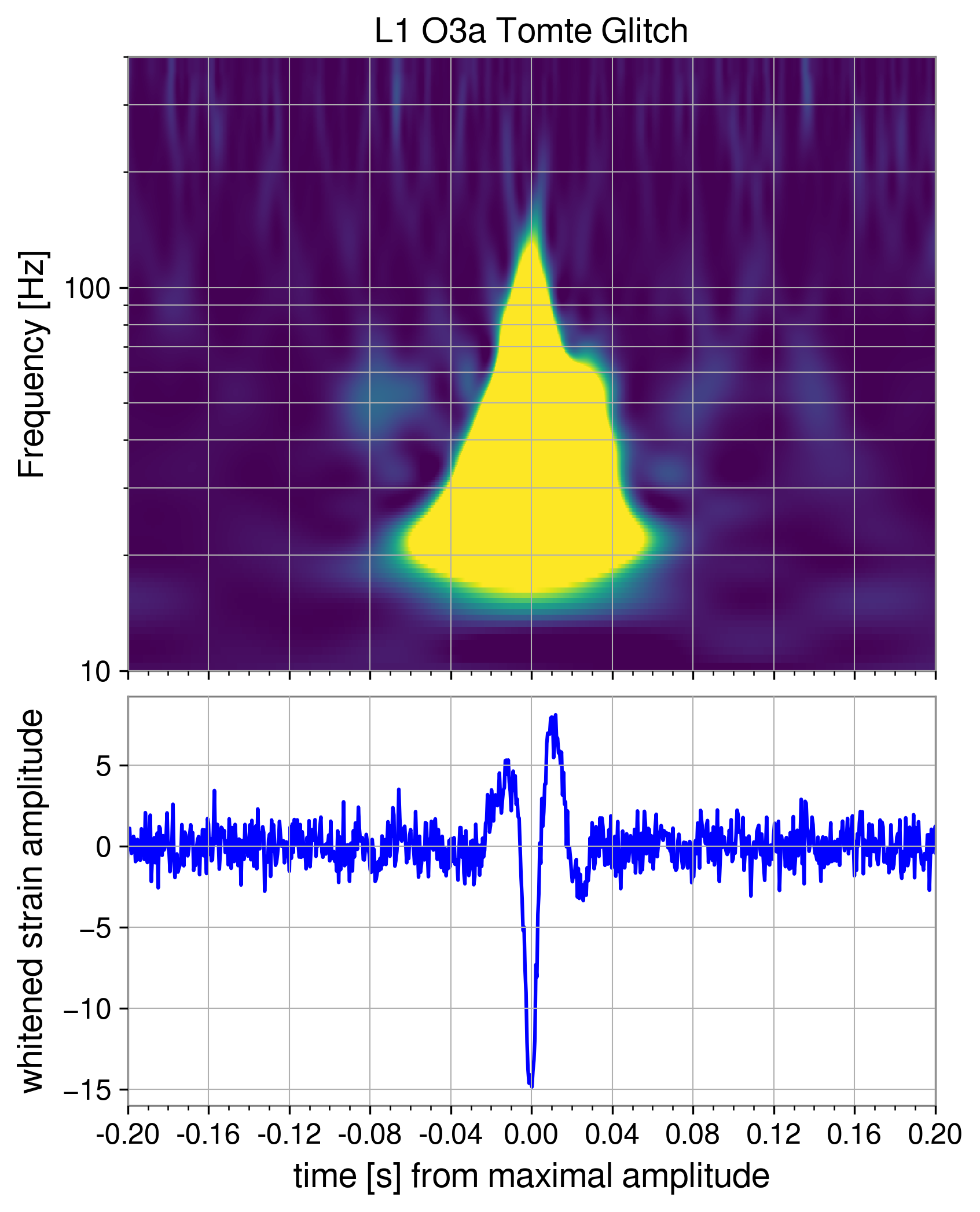}
    \end{minipage}

\caption{Typical loud blip (above) and tomte (below) glitches from the test set for Livingston in O3a, to demonstrate morphology. Q-scans indicate power as color in time-frequency pixels, and the timeseries (below in blue) shows additional morphology. Note that timescales and frequency ranges plotted vary. Blips are sometimes shorter than 5ms, where tomtes can last over 100ms.} \label{fig:blip_tomte}

\end{figure}
\FloatBarrier

\subsection{The Transient Glitch Background}

The characteristics of the noise background are well covered in \cite{detchar, gspy1, gspy2021, noise_guide}. The morphologies of a typical `blip' and `tomte' glitch are explored in Figure \ref{fig:blip_tomte}. \NewChange{Blip glitches are short, at around 5-10ms, while Tomtes are typically ~100ms long.} To properly mitigate these glitches we examine their morphology as they appear to searches, \textit{after} any whitening and postprocessing. Physically, it is possible that the glitches are a very brief DC offset that appears in the strain channel, the result of either a single physical perturbation to some component of the detector or the result of a digital error. We will have to consider the additional morphology of finite impulse response whitening filters as being part of the glitch, since the searches must also contend with these features.

\FloatBarrier
While GravitySpy examines q-transforms \cite{q-transf} of glitches, \NewChange{we train on the frequency series of glitches}. There is a loss of phase information and direction of amplitude in q-transforms, which record only power for each time-frequency pixel. This may be important for future efforts in distinguishing auxiliary witnesses for these glitches, since a preferential directional perturbation to a part of the detector could show up as a bias in amplitude (positive or negative) in the strain channel for a certain detector and glitch type. We have yet to determine if this is a bias introduced in GravitySpy's curation of the highest confidence and loudest glitches, or if this extends to the large number of lower confidence glitches as well, but we see a vast majority of confident, loud L1 O3a tomtes with negative amplitudes. Other detectors and glitch types exhibit a certain ``glitch-signature'' in amplitude bias, sometimes across multiple observing runs. 

\subsection{The \textit{glitschen} Model}

\par In the \textit{glitschen} parametric glitch mitigation model, we employ probabilistic principal component analysis (PPCA). This is a simple and effective way for us to decompose a frequency domain signal into a set of Gaussian distributed latent variables. It is frequently used as a  dimensionality reduction tool, making problems in many areas of data science more tractable. There are many ready-made principal component analysis (PCA) implementations available. We found it most transparent and effective to write our own PPCA implementation, closely following the original PPCA model \cite{microsoft_PPCA}. This enabled us to find a fast and computationally cheap way to analytically maximize our likelihood. PPCA differs from PCA in that it includes a Gaussian noise term. 

We employ an isotropic Gaussian noise model:
\begin{equation}
\mathcal{N}(\bm{0}, \sigma^{2}\bm{I})
\end{equation}
with a d-dimensional observation vector, $\bm{\tilde{d}}$:
\begin{equation}
\bm{\tilde{d}}|\bm{Z_{train}} \sim \mathcal{N}(\bm{WZ_{train}}+\bm{\mu}, \sigma^{2}\bm{I})
\end{equation}
We assume the marginal distribution $\bm{Z_{train}} \sim \mathcal{N}(\bm{0},\bm{I})$ over q latent variables of the training set, and $\bm{W}$ has size $d \times q$, containing q training eigenvectors. We recover normal PCA in the limit of $\sigma \xrightarrow{} 0$. 
We can marginalize over the latent variables to obtain a distribution for $\bm{\tilde{d}}$:
\begin{equation}
\bm{\tilde{d}} \sim \mathcal{N}(\bm{\mu,C})
\end{equation}
where $\bm{C} = \bm{WW}^{T} + \sigma^2 \bm{I}$ is the covariance model for the observed data, with dimension $d \times d$. In our case this data is frequency-series data.  With $N$ training glitches, our log-likelihood for the entire model and all our observed (training) data is then:
\begin{equation}
\ln\mathcal{L}_{training} = -\frac{N}{2} \bigg{[} d \ln(2\pi) + \ln |\bm{C}| + tr(\bm{C}^{-1}\bm{S})\bigg{]}
\end{equation}
with the sample covariance matrix of the observations, $\bm{S}$:
\begin{equation}
\bm{S} = \frac{1}{N}\sum_{n=1}^{N}(\bm{\tilde{d}}_{n}-\bm{\mu})(\bm{\tilde{d}}_{n}-\bm{\mu})^{T}
\end{equation}
This likelihood is often maximized iteratively, and many packaged implementations of PPCA find $W$ in this way \cite{dillon2017tensorflow}. However we find the global maximum of the likelihood using an analytical method detailed in \cite{microsoft_PPCA}. 
\par
\begin{figure} \label{fig:eigplot}
    \includegraphics[width=0.8\linewidth]{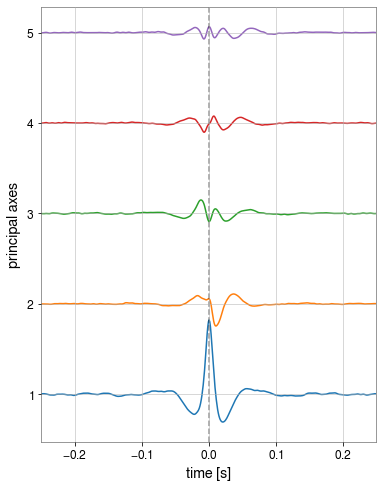}
    \caption{L1 O3a tomte glitch model eigenvectors. Increasing weight from top to bottom.}
\end{figure}

 Later, we use this likelihood, with an Occam's penalty accounting for the effective degrees of freedom in the model, to find the optimal number of components, q, to use.
Performing an eigenvalue decomposition on $\bm{S}$, the sample covariance matrix of the observations, we obtain the $d \times q$ matrix $\bm{U}_{q}$ containing q principal eigenvectors (or ``eigenglitches") of $\bm{S}$, and the $q \times q$ diagonal matrix $\bm{\Lambda}_{q}$ with cooresponding eigenvalues. All eigenvalue decompositions and matrix inversions are conveniently handled by an open-source computer algebra library with numpy\cite{van_der_Walt_2011}. The likelihood is maximized when:
\begin{equation}
\bm{W} = \bm{W}_{ML} = \bm{U}_{q}(\bm{\Lambda}_{q}-\sigma^{2}\bm{I})^{1/2}
\end{equation}
Going forward, we can consider $\bm{W}=\bm{W}_{ML}$ to always contain the maximum likelihood (ML) eigenvectors. See an example time-domain representation in Figure \ref{fig:eigplot}.
\par 
In order to obtain a projection of a new observation vector, $\bm{\tilde{d}}_{obs}$, onto the latent variables we use Bayes rule to get from $\bm{\tilde{d}}|\bm{Z_{train}}$ to 
\begin{equation}
\bm{Z_{train}}|\bm{\tilde{d}_{obs}} \sim \mathcal{N}(\bm{M}^{-1}\bm{W}^{T}(\bm{\tilde{d}_{obs}}-\bm{\mu}),\sigma^{2}\bm{M}^{-1})
\end{equation}
where $\bm{M} = \bm{W}^{T}\bm{W} + \sigma^{2}\bm{I}$, with dimensions $q \times q$. This allows us to perform reconstructions of suspected glitches using a trained model. We define
\begin{equation}
\bm{Z_{rec}}\equiv \bm{M}^{-1} \bm{W}^{T}  (\bm{\tilde{d}_{obs}}-\bm{\mu})
\end{equation}
as the set of $q$ latent optimal (i.e., maximum likelihood) reconstruction weights up to an arbitrary rotation matrix.

\FloatBarrier

We can obtain a reconstruction with: 

\begin{equation}
\bm{\tilde{g}}_{rec} = \bm{WZ_{rec}} + \mu 
\end{equation}

To evaluate the quality of our reconstruction given the data, we employ the standard Gaussian noise likelihood, identical to that used by CBC searches and PE, and specifically \textsc{Bilby} \cite{bilby}. We define the standard noise weighted inner product of any two frequency-series vectors, $\bm{a}$ and $\bm{b}$:
\begin{equation}
(\bm{\tilde{a}}|\bm{\tilde{b}}) = 2 \int^{\infty}_{0}\frac{\tilde{a}(f)\tilde{b}^{*}(f) + \tilde{a}^{*}(f)\tilde{b}(f)}{S_{n}(f)}df
\end{equation}
$S_{n}=\sigma^{2}$ is the noise power spectral density (PSD), and $\sigma$ is the amplitude spectral density (ASD)\cite{noise_guide}. In practice, the noise term can be taken to be 1, because the model is trained on whitened data. 
When we assume stationary, Gaussian noise that is uncorrelated between detectors, our reconstruction log-likelihood becomes:
\begin{equation}
\ln\mathcal{L}_{rec} = -\frac{1}{2}\sum_{k}\bigg\{\frac{|\bm{\tilde{d}}_{obs,k}-\bm{\tilde{g}}(\theta)_{rec,k}|^{2}}{S_{k}} + \ln (2\pi S_{k})\bigg\}
\end{equation}
where k is the frequency bin index $\bm{\tilde{g}}(\theta)_{rec,k}$ is the frequency domain reconstruction with PPCA parameters $\theta$. With this inner product and likelihood we can compare our model's reconstruction of an event, after training on a certain glitch class, with the likelihood of the astrophysical hypothesis. We can select $q$ based on an Occam's penalty, or we can try to replicate the number of effective free parameters in the CBC model to give equal flexibility.

\subsection{Implementation and Performance}
\subsubsection{Selection of Training Data}
\par We curate glitches classified by GravitySpy \cite{gspy1} with high `confidence', where the score ranges from (0,1). Note that confidence is not a normalized probability, but instead reflects the certainty of classification by the convolutional neural network used. We utilize the newest, LVK-internal version of the GravitySpy model, which has the benefit of training on data from all of O3. Publicly available glitch and event data can be obtained from the Gravitational Wave Open Science Center (GWOSC) \cite{gwosc_o1o2}. This analysis was completed using an older version of the calibrated data: the HOFT\_C00 strain data frame within the GDS-CALIB\_STRAIN\_CLEAN channel. Note that some ($<$ 1\%) of the glitches used in training are outside of ``science mode" times. 

All glitches used first must clear our confidence cutoff\NewChange{(0.95-1, depending on type, detector, and epoch), and are then sorted by SNR}. Lower SNR glitches can contaminate the model with more unrelated noise features. \NewChange{As such, we have kept a high SNR threshold for inclusion in training(dependent on type, detecter, and epoch), where we use the 1500-2000 loudest glitches.} It is more productive to limit the set to ``golden" examples curated by GravitySpy, even if the glitch or event in the run segment has low SNR, since we believe quiet and loud glitches (~5-50 SNR) exhibit similar morphology, based on our exploration of the data. 
\subsubsection{Preprocessing and Training}
\par To train our model, we whiten with an ASD calculated from between 16 and 128 seconds of data, depending on the glitch-type in question. Because we are concerned with the low-frequency content of glitches (in the range of astrophysical searches) all data is downsampled to 2048 Hz, and then for certain glitchtypes we further bandpass training data to aid in reconstruction efficiency. For Tomte glitches, which have a peak frequency around 50-60 Hz, a 10-128 Hz bandpass to the training data ensures we are not overfitting noise outside the glitch time, but still recover more than 99\% of the SNR from more than 99\% of training glitches. We find that 0.5s training window is always adequate for tomtes, with typical duration ~0.1s. For blips, peak frequencies are typically 500-1000 Hz, so we obtain similar recovered SNR by bandpassing from 10-1024 Hz. We find that a 0.1s window is almost always adequate for blips (allowing one full cycle at 10 Hz). Blips are shorter in duration (almost always shorter than 30ms). For run segments on test glitches and marginal/glitch-like events we keep data in 10-2048 Hz, retaining higher frequency noise. All training examples are centered on the peak amplitude time sample. All preprocessing is performed using open-source libraries including \textsc{NumPy} \cite{van_der_Walt_2011} and \textsc{GWPy} \cite{gwpy}.

\subsubsection{Performance}
The model is easily run and bench-marked on a laptop with six cores. The training process takes less than 1 second for ~2000 glitches. Maximum likelihood reconstruction takes 1ms-1$\mu$s depending on how much leeway in center time we allow. Sampling proceeds quickly, giving 10,000 independent samples of the posterior distribution in about 5 minutes, depending on the glitch. 

By weighing the likelihood against an Occam's penalty, we can ensure our model has the appropriate number of dimensions ($q$) and is not over fitting. We employ a Laplace approximation to the marginal likelihood \cite{automaticchoiceMinka00}, along with the Bayesian information criterion, described further in appendix A, to choose the optimal number of eigenvectors for calculating the residuals of the test sets, in the next section. To roughly match the degrees-of-freedom (per-detector) of the CBC model, we employ $q=5$ in all sampled cases.

\section{Results}

\subsection{Testing with Maximum Likelihood Reconstruction}

We reserve 10\% of glitches for testing (the model has never encountered these examples), and to evaluate the performance of our model we examine residuals after maximum likelihood reconstruction and subtraction, as seen in Figure 3 histograms. This demonstrates the efficacy of the model in mitigating an entire class of glitches. Test sets shown include 100-200 glitches. We will soon extend this to cleaning entire search backgrounds, and attempt re-ranking of CBC searches. 

We plot residuals after glitch subtraction in the frequency domain. They obey a Gaussian distribution after perfect glitch cleaning under the hypothesis of stationary, uncorrelated noise. Cleaning models are trained with the automatic choice of dimensionality via Laplace approximation (described in Appendix A): (15, 2, 9, 8), for H1 blips(truncated from 23 to 15), H1 tomtes, L1 blips, L1 tomtes, respectively.

\begin{figure}

\begin{minipage}{.48\textwidth}
    \includegraphics[width=\linewidth]{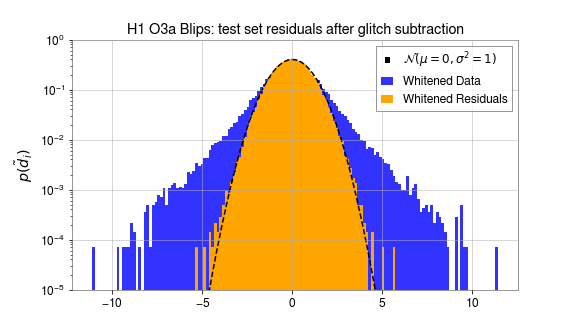}
    \end{minipage}
    \begin{minipage}{.48\textwidth}
    \includegraphics[width=\linewidth]{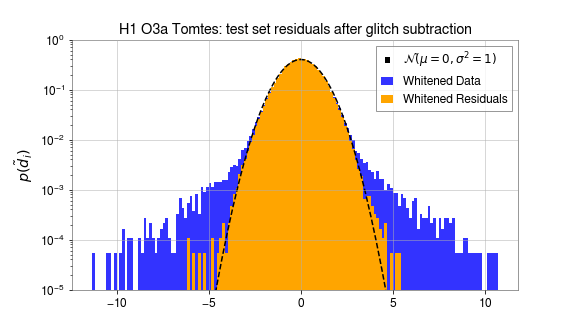}
    \end{minipage}
    \begin{minipage}{.48\textwidth}
    \includegraphics[width=\linewidth]{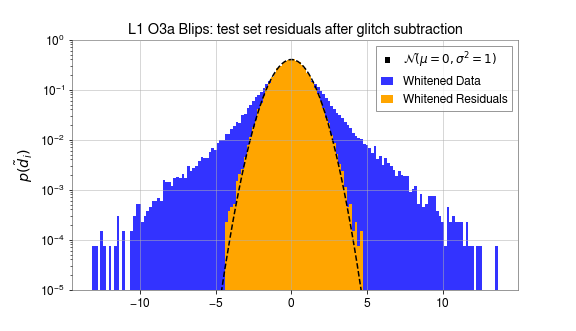}
    \end{minipage}
    \begin{minipage}{.48\textwidth}
    \includegraphics[width=\linewidth]{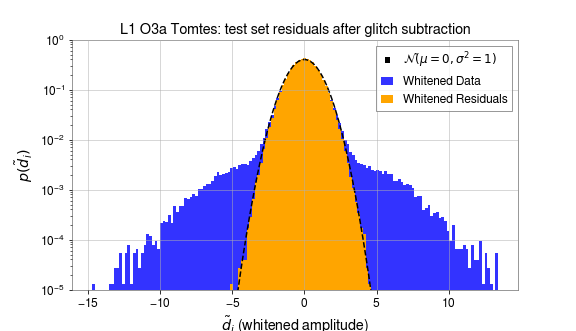}
    \end{minipage}
\caption{Frequency-domain residuals after subtraction from the test set (10\%) reserved from each glitch type, detector, and epoch. The bins are scaled such that the lowest visible represent single samples from single glitches. Note that extremal samples are louder in Livingston. It has been observed that with greater sensitivity and range transient glitches become louder as well \cite{gspy2021}.} \label{fig:whitened}
\end{figure}

The binning in Figure \ref{fig:whitened} extends down to single samples from single test glitches, showing that for all classes and detectors our results are consistent with Gaussian noise. The performance is somewhat higher for tomte glitches, mainly due to the greater homogeneity in their morphology compared to blips. Tomtes in Livingston were 10-20 times more prevalent than in Hanford \cite{gspy2021}. This has been partly attributed to Livingston operating at greater sensitivity than Hanford during O3a, but may also be to to unknown environmental factors. It is observed that blips and tomtes are louder at higher sensitivity. Higher SNR and greater numbers allow for better modelling, but show the increasing importance of mitigation as sensitivity improves in future observing runs. 

\begin{figure*}

\begin{minipage}{0.8\textwidth}
    \includegraphics[width=\linewidth]{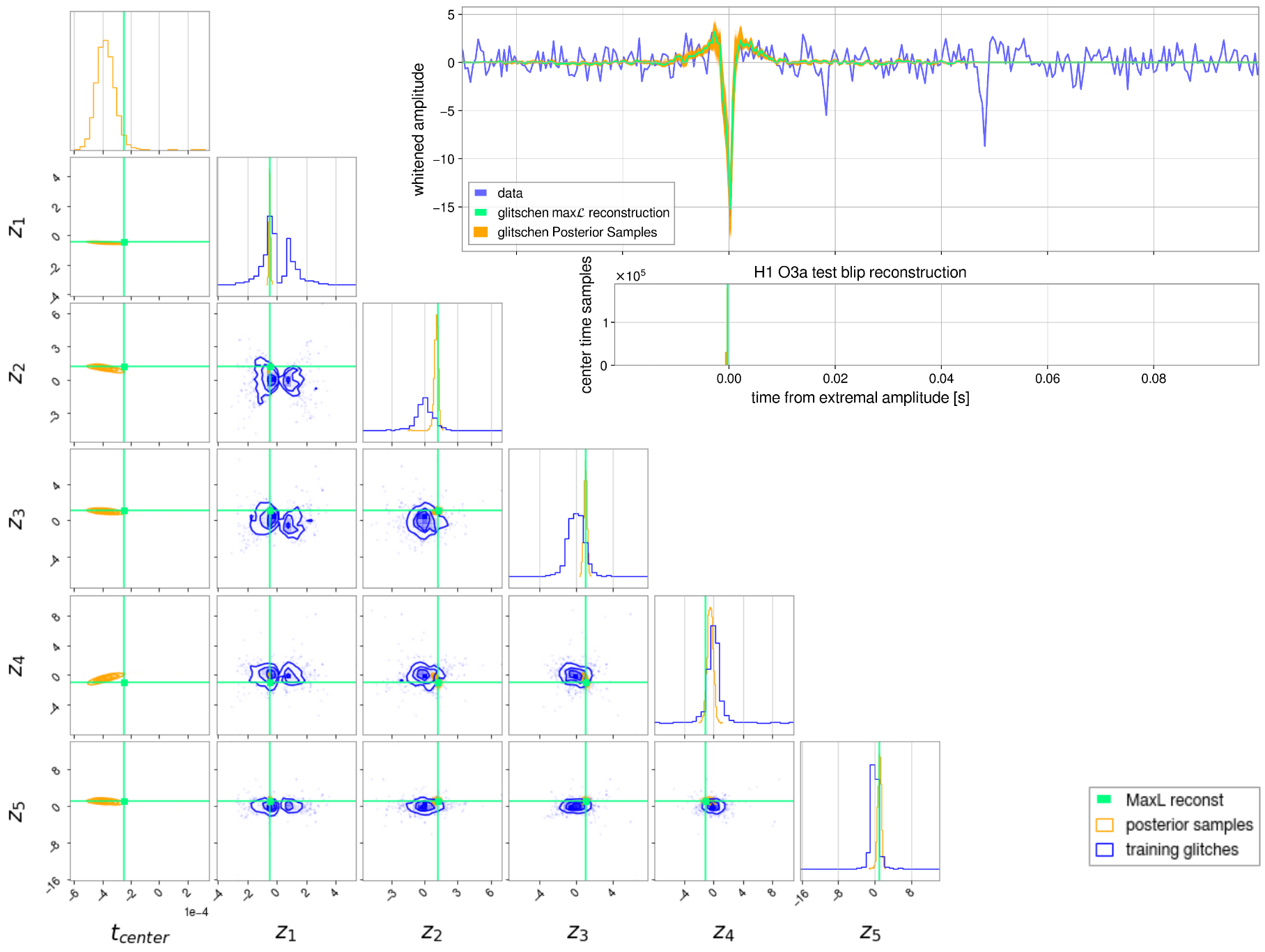}
    \end{minipage}

\caption{H1 O3a test blip: full posterior estimation. Note the repeating blips afterward. This example shows the tendency of the sampler to converge on the loudest glitch available. The histogram of center time samples shows high certainty (just below the timeseries reconstruction). In the corner plot for the latent space weights (z$_{1}$-z$_{5}$), we see that this test set glitch is typical of the class.}\label{fig:test_blip}

\begin{minipage}{0.8\textwidth}
    \includegraphics[width=\linewidth]{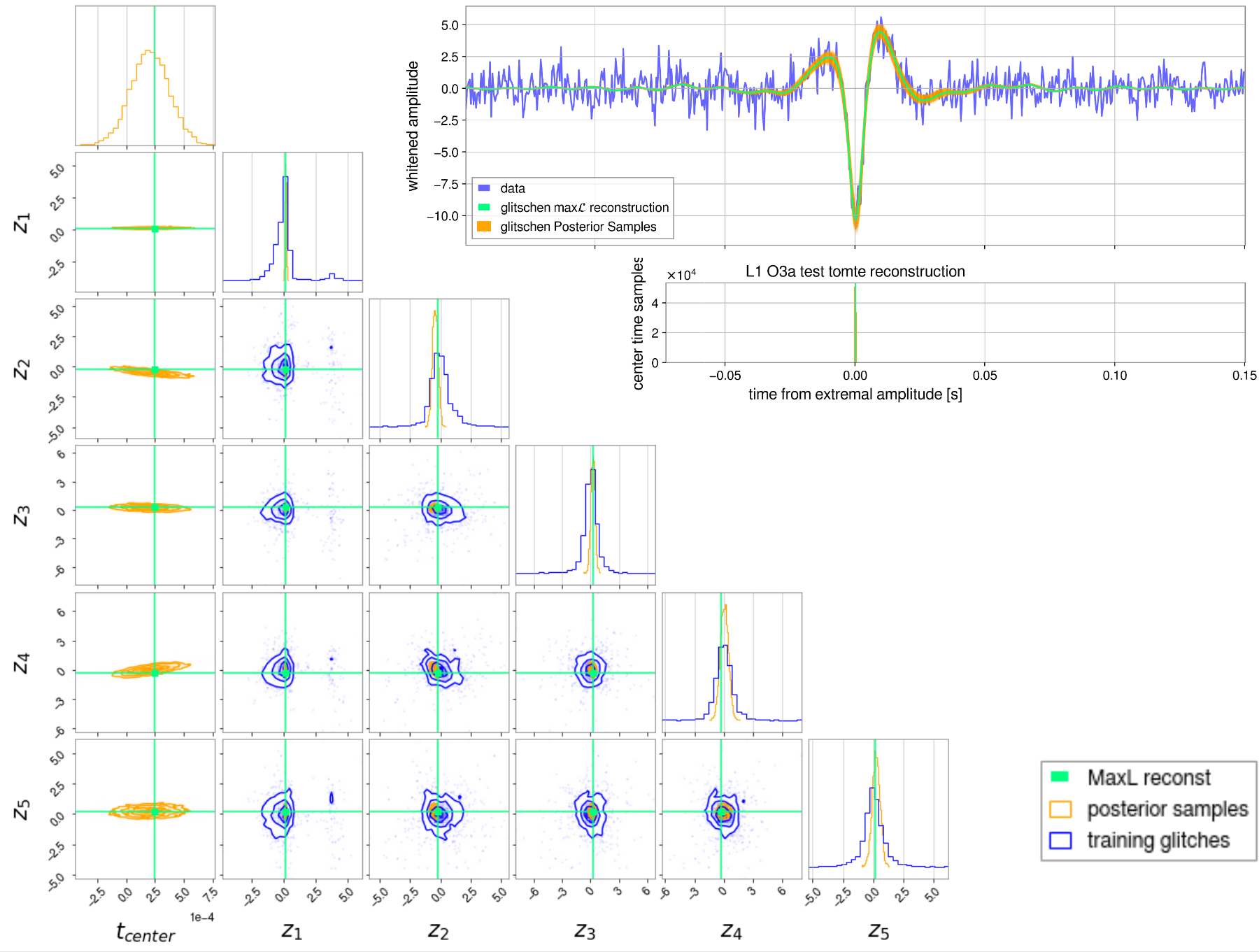}
    \end{minipage}

\caption{L1 O3a test tomte: full posterior estimation. This is a very typical tomte glitch, with all walkers converging on the same center time, low uncertainties in the timeseries reconstruction, and the posterior distribution aligning well with the distributions on the training set latent weights.} \label{fig:test_tomte}
\end{figure*}
\FloatBarrier

\subsection{Sampling} 
We employ two well developed MCMC toolkits, \textsc{emcee} \cite{emcee}, and \textsc{kombine} \cite{kombine}, to perform a full Bayesian posterior estimation of our reconstruction. By allowing the center time of the hypothesized glitch to vary, we sample in q+1 dimensions. A priori, we assume glitches are equally likely at any time, and thus adopt a uniform prior in center time. The localization of the samples in center time is a good indicator of how glitchlike the morphology of the test signal is. To aid in the efficiency of sampling, we initialize walkers in a Gaussian around the suspected glitch time. In the $q$ PPCA weights, we use a less restrictive wide Gaussian prior, or alternately a highly-informed KDE (kernel density estimate) prior based on the entire training set's maximum likelihood weights. The latter is generally more restrictive and can limit the flexibility of the sampler to fit more general morphologies, which in some cases may be ideal, and in others can to be adjusted. For all MCMC sampled example glitches and CBC comparisons in the paper, we employ $q=5$. 

In Figures \ref{fig:test_blip} and \ref{fig:test_tomte} we demonstrate the results of sampling on a test blip in Hanford, and a test tomte in Livingston. The blip was chosen specifically due to it's proximity to further repeating blips. The sampler converges easily on the loudest glitch-like event in the run-segment. 

\FloatBarrier

\subsection{Signal Safety Testing}
To establish the model's capability of distinguishing glitch from astrophysical signal, we test if it remains flexible enough to fit different glitch morphologies while being (appropriately) unable to reconstruct and subtract an astrophysical signal. We run our model on a selection of high-mass, short duration BBH signals from GWTC-2  \cite{gwtc2}, acquiring data from the Gravitational Wave Open Science Center(GWOSC)\cite{gwosc_o1o2}. Specifically, we choose events with high detector frame chirp masses (and by extension, short template durations), and FAR (False Alarm Rate)$<10^{-3}/yr$. Being the confirmed astrophysical signals with morphology closest to short transient glitches such as blips and especially tomtes, these provide a good opportunity to confuse the model. We quote maximum likelihood single detector SNR values for the CBC and alternately the glitch hypotheses in Table \ref{table:1}. Quoted durations are the template duration for the preferred trigger from low latency detection. Two events had H1 offline at the trigger time but were included for their glitchlike morphology. We anticipate that an effective glitch model may be instrumental in vetting single-detector events in the future. 

For a more rigorous comparison between the CBC and glitch models we employ a full posterior estimation framework and the Deviance Information Criterion (DIC). This is a variance based approach. The DIC is given by:
    \begin{equation}
    DIC = D(\Bar{\theta}) + \overline{\mathrm{var}(D(\theta))}
    \end{equation}
Where the deviance, D, is: $D(\theta) = -2\ln(p(d|\theta))$ with posterior distribution p, data y, and parameters $\theta$.
Given the log likelihoods from samples obtained using both the Glitchen model and a CBC PE run we see that the glitch hypothesis is heavily disfavored for all events tested. These comparisons appear in Table II, where a lower DIC value indicates a better model for the observed data.

\begin{table*}[]
\centering
\begin{tabular}{|c|c|c|c|c|c|c|c|c|c|}
\hline

 \multicolumn{3}{|c|}{Event Information} & & \multicolumn{6}{|c|}{Matched-Filter SNR} \\
\hline    
Event Name & $\mathcal{M}_\mathrm{det}$, $M_{\odot}$ & duration(s) && CBC H1 & Tomte H1 & Blip H1 & CBC L1 & Tomte L1 & Blip L1 \\ [0.5ex]
\hline

    GW190521 & $114.8^{+15.2}_{-17.6}$ & 0.15 && 7.87 & 4.11 & 3.21 & 12.38 & 5.93 & 4.06 \\
    GW190602\_175927 & $72.9^{+10.8}_{-13.7}$ & 0.22 && 6.56 & 3.60 & 3.55 & 11.02 & 4.43 & 4.88 \\
    GW190706\_222641 & $75.1^{+11.0}_{-17.5}$ & 0.15 && 9.07 & 4.91 & 4.22 & 9.18 & 3.92 & 3.93 \\
    GW190519\_153544 & $65.1^{+7.7}_{-10.3}$ & 0.17 && 9.50 & 4.42 & 4.76 & 11.85 & 5.42 & 4.25 \\
    GW190620\_030421 & $57.5^{+9.0}_{-11.2}$ & 2.3 && (offline) & - & - & 11.70 & 3.78 & 4.63 \\
    GW190910\_112807 & $43.9^{+4.6}_{-3.6}$ & 1.8 && (offline) & - & - & 13.86 & 6.29 & 4.26 \\
    GW190521\_074359 & $39.8^{+2.2}_{-3.0}$ & 0.24 && 12.67 & 5.85 & 6.30  & 22.68 & 8.83 & 7.24 \\
\hline
\end{tabular}
\caption{Selecting short duration, heavy BBH mergers we provide an important test for the model, which should give lower SNRs than the CBC model. Events are in order of detector frame chirp mass ($\mathcal{M}_\mathrm{det}$, $M_{\odot}$). For all of these events we see lower SNRs by a factor of 2-3, whereas we expect to recover nearly all of the SNR in confirmed glitches. CBC parameter estimation results from \cite{gwtc2}.}
\label{table:1}
\vspace*{0.5cm}
\centering
\begin{tabular}{|c|c|c|c|c|c|c|c|}
\hline
Event Name & \multicolumn{6}{c}{Deviance Information Criterion (DIC)} & \\ 
\hline
 & CBC H1 & Tomte H1 & Blip H1 && CBC L1 & Tomte L1 & Blip L1 \\  [0.5ex]
\hline

    GW190521 & -54.6 & -5.4 & 8.7 && -130.3 & -26.7 & 9.94 \\
    GW190602\_175927 & -37.3 & 10.2 & 8.7 && -105.6 & -5.6 & 12.0 \\
    GW190706\_222641 & -71.0 & 13.2 & 9.6 && -74.0 & 14.8 & 8.1\\
    GW190519\_153544 & -87.7 & 18.4 & -18.6 && -145.2 & 25.1 & 13.1 \\
    GW190620\_030421 & - & - & - && -133.9 & 17.5 & 12.5 \\
    GW190910\_112807 & - & - & - && -190.3 & -28.5 & 17.7 \\
    GW190521\_074359 & -152.8 & 34.0 & 36.4 && -494.5 & -67.1 & -45.3 \\
\hline
\end{tabular}
\caption{Running samplers on these events, we obtain the DIC from our distributions of log likelihoods. The deviance information criterion (DIC) favors models with a lower value. The CBC model is highly preferred to the glitch model in all cases, indicating that pass the signal safety test.}
\label{table:2}
\end{table*}

\subsection{GW190521: testing our model's limits with the most massive (and glitch-like) confident O3a event}
GW190521 is the highest mass ($142^{+28}_{-16}
M_{\odot}$) and shortest duration (~0.1s) CBC event for which we have strong evidence \cite{190521_discovery}. Being a loud triple detector event, it is confidently of astrophysical origin. But for us, it offers a unique opportunity to test our model, since it exhibits signal morphology which is the most ``glitch-like" of all high significance astrophysical events. It spent only the last 4 cycles of its inspiral in the sensitive band of the detector, peaking at 60Hz. Tomte glitches look very similar. 

Critically, any model for tomtes, at bare minimum, must not be confused by such an event. Because the aim of improved glitch mitigation is to open up this high mass region of parameter space, this is precisely the kind of test we need to pass. Here we demonstrate our full posterior estimation framework on GW190521, and by extension, our ability to distinguish glitch-like astrophysical events from glitches by comparing our glitch hypothesis results with the astrophysical hypothesis results.

\begin{figure*}

\begin{minipage}{0.84\textwidth}
    \includegraphics[width=\linewidth]{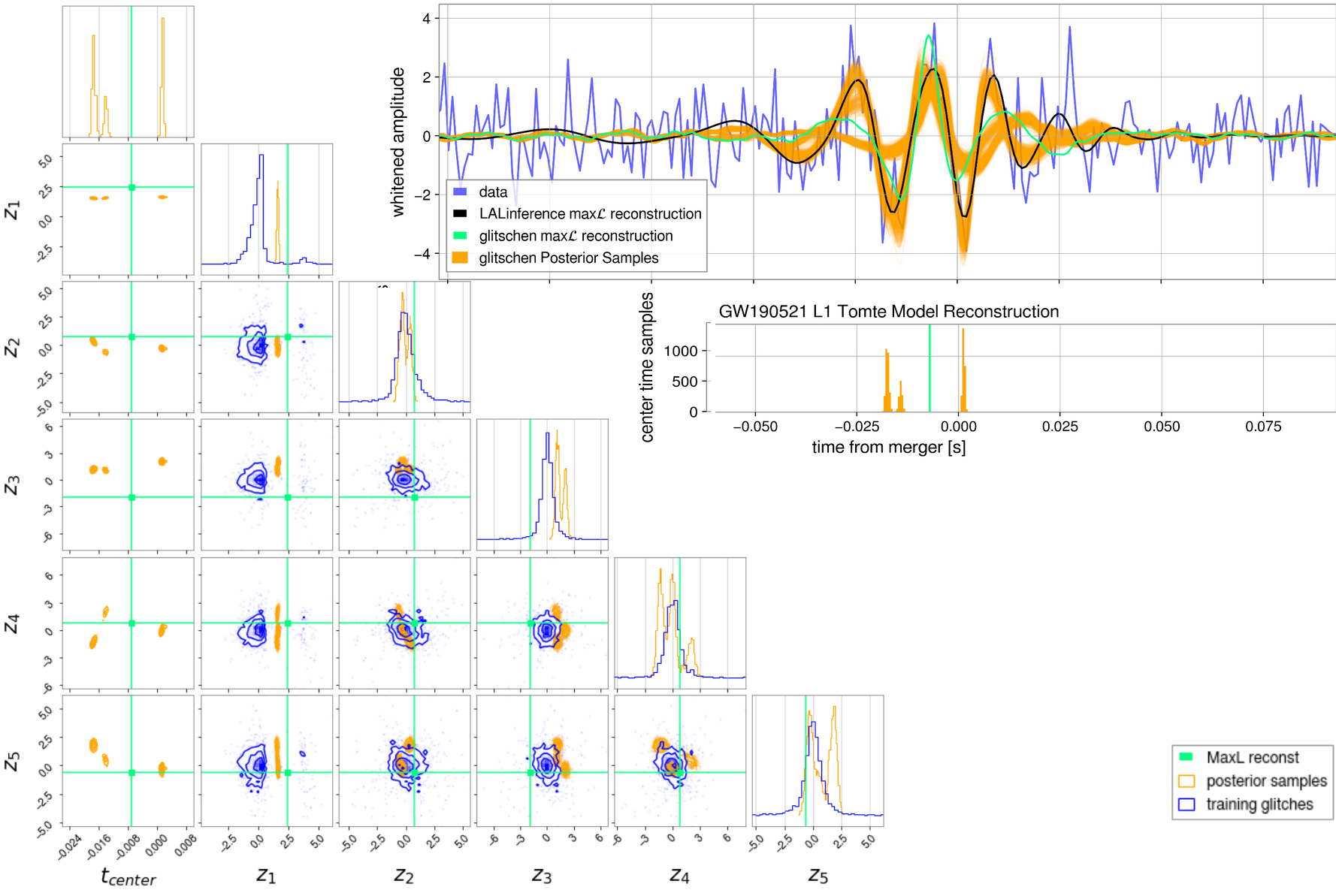}
    \end{minipage}
    
\caption{GW190521 Full Posterior Estimation, L1. The distribution in the center time is multi-modal, indicating that the glitch model fails to capture the full morphology of the signal (\textsc{LALInference} maxL in black), no matter where it is placed. The reconstruction features high uncertainty (samples in orange), and the posterior distribution in the latent variables lies outside the training set of glitches. All of this indicates that the model has failed to reconstruct GW190521 as a glitch, as expected.} \label{fig:l1190521}
\end{figure*}

\begin{figure*}

\begin{minipage}{0.85\textwidth}
    \includegraphics[width=\linewidth]{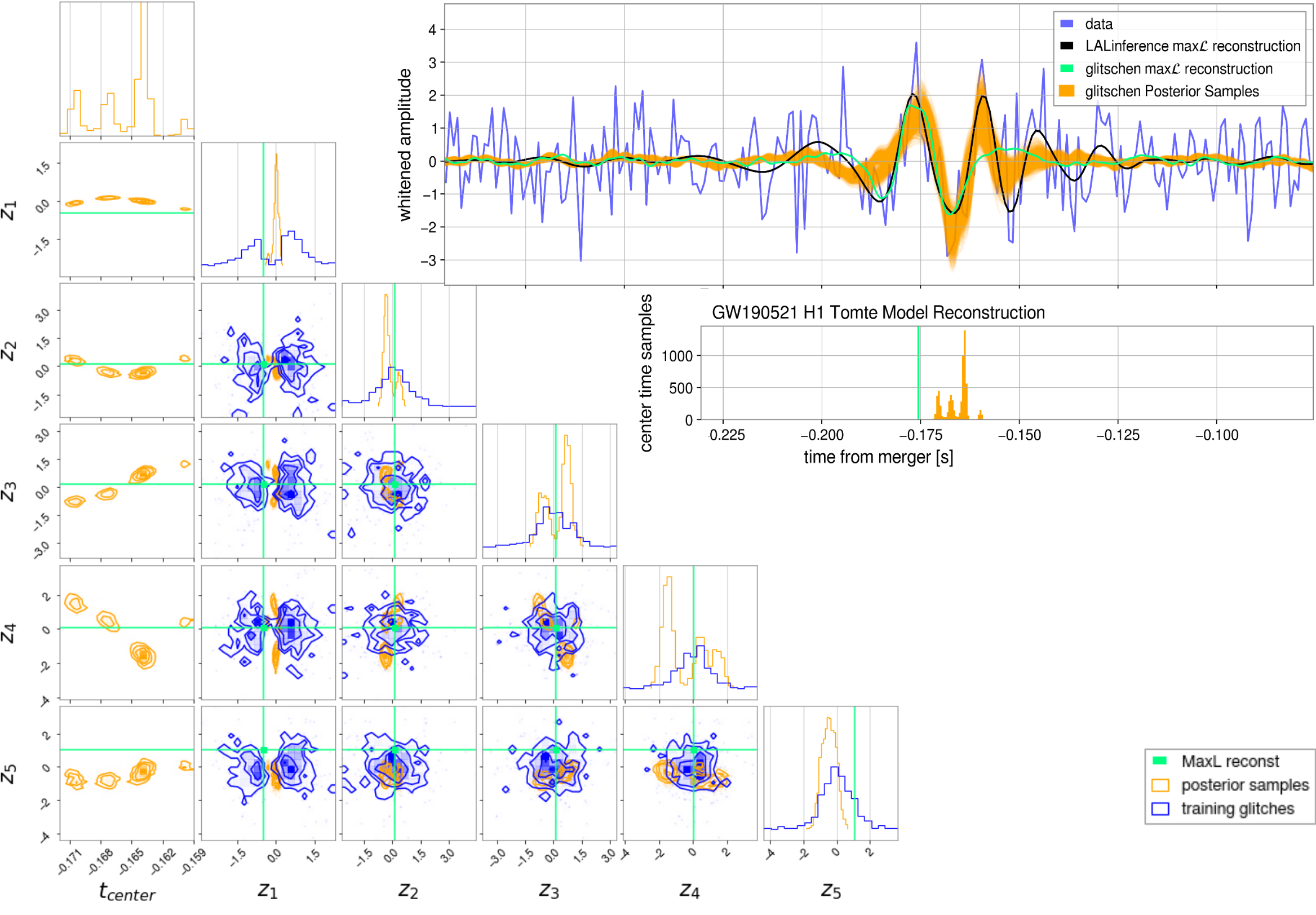}
    \end{minipage}

\caption{GW190521 Full Posterior Estimation, H1} \label{fig:h1190521}
\end{figure*}

In both L1 and H1 (Figures \ref{fig:l1190521} and \ref{fig:h1190521}, respectively), we see that the glitch model (MaxL glitch reconstruction in green) is unable to fully capture the signal morphology, with the MaxL CBC reconstruction in black, no matter where it is placed. It remains multi-modal in center time, and an outlier in most of the training set weights, indicating that this is a poor fit to the data, as we expect. High uncertainty is seen in the broadness of the posterior reconstructions, in orange on the timeseries plots. See Tables I and II for a more quantitative comparison of the glitch and CBC hypothesis, for this and other events in O3a.

\section{Conclusion and Future Work}

We have introduced a PPCA-based approach to modeling transient noise in gravitational wave detector data, implemented in the open-sourced \textit{glitschen} package, publicly available here: \footnote{https://git.ligo.org/jonathan.merritt/glitschen}. We welcome collaborative development, testing, and feedback. 

For both `blip' and `tomte' glitches -- some of the most impactful for \NewChange{BBH} searches in O3 -- we have demonstrated the effectiveness of the model for glitch subtraction, as well as for Bayesian model comparisons with astrophysical signal models.

In future work we will explore the use of clustering algorithms in PPCA space for glitch classification and sub-classification.  We will test the effectiveness of the model in reducing the background for compact binary searches. We will also integrate our model into the \textsc{Bilby} \cite{bilby} parameter estimation code, where composite signal and noise models will allow us to marginalize over glitch morphology when glitches are coincident with astrophysical signals. 

We tested our model on high-mass events from O3a, but in the future we will extend this testing to simulations in the high-mass \textit{and high mass-ratio} region of parameter space, where discoveries are still to be made and distinguishing astrophysical events from noise is even more difficult.

Because burst searches also trigger on glitches, we plan to test our model in this regime. Searches for cosmic string cusps, supernova templates, and all agnosticly un-modeled sources could radically change the field, but only if we can work on the serious blind-spots in our searches. We have already began an injection campaign with cosmic string templates in the parameter space contaminated by blip glitches to determine our ability to differentiate signal from glitch in this context.  \NewChange{We plan to extend the use of our model beyond Blips and Tomtes, but because these are the most impactful for BBH searches, they remain the first and most important testing ground.}

With more accurate models of glitches, we can improve the detectability and significance of gravitational wave events of all kinds.

\section{Acknowledgements}

Here we thank the GravitySpy team, the detector characterization working group, and everyone who made this effort possible. We're grateful to the CBC, Detchar, and Parameter Estimation groups for their hard work building powerful software tools and tutorials and organizing workshops that help make LIGO data analysis work more accessible. 

The authors are grateful for computational resources provided by the LIGO Lab (CIT, LHO, LLO)  and supported by National Science Foundation Grants PHY-0757058 and PHY-0823459. This material is based upon work supported in part by the National Science Foundation under Grant PHY-2110636. We are grateful for the following open-source software tools: \textsc{NumPy}~\citep{van_der_Walt_2011}, \textsc{SciPy}~\citep{2020SciPy-NMeth}, \textsc{GWpy}~\citep{gwpy}, \textsc{Matplotlib}~\citep{Hunter:2007}, \textsc{bilby}~\citep{bilby}.

\bibliographystyle{apsrev4-1}
\bibliography{bibliography}

\appendix
\section{Optimal Choice of the Model Dimensionality}

\begin{figure}[H]

\begin{minipage}{.45\textwidth}
    \includegraphics[width=\linewidth]{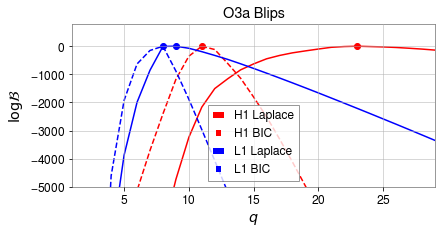}
    \end{minipage}
    \begin{minipage}{.45\textwidth}
    \includegraphics[width=\linewidth]{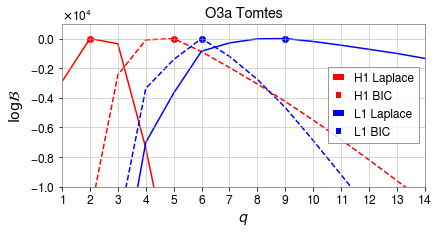}
    \end{minipage}

\caption{The relative Bayes factors as a function of dimensionality, q, for each detector and glitch type in the analysis. The peak of these curves allow for an automatic choice of dimensionality that avoids over-fitting.} \label{fig:dimensionality}
\end{figure}

To avoid over or under fitting we can use various metrics to find the optimal number of PPCA eigenvectors to employ for each glitch type, detector, and observing run. We tried a crude method: fraction of recovered SNR in test set glitches. If we recover .99 of the known glitch SNR then any gains added with additional dimensions are giving diminishing returns. However this cutoff point is somewhat arbitrary. Instead, balancing an Occam's penalty against the model's training set likelihood is a much more rigorous approach. We employed several methods, including the Akaike information criterion (AIC), the Bayesian information criterion (BIC), and the Laplace approximation to the marginal model log-likelihood, following the method in \cite{automaticchoiceMinka00}. By maximizing these metrics we can use the optimal level of model complexity.  To arrive at the Laplace approximation, we apply an uninformative conjugate prior on the model parameters and marginalize over everything but q, the PPCA dimensionality. The marginal log-likelihood values are estimates of the model evidence and the ratio of these for different q can be taken as Bayes factors, so far as the Laplace approximation is accurate, which we show in Figure \ref{fig:dimensionality}. 

\end{document}